\documentclass[aps,prl,reprint,twocolumn,floatfix,superscriptaddress,longbibliography,groupedaddress]{revtex4-2}

\usepackage{standalone}
\usepackage{amsthm}
\usepackage{lipsum}
\usepackage{amsfonts, amsmath, amssymb}
\usepackage{graphicx}
\usepackage{dcolumn}
\usepackage{bm}
\usepackage{dsfont}
\usepackage[normalem]{ulem}
\usepackage{color}
\usepackage[utf8]{inputenc}
\usepackage[colorlinks,citecolor=blue,linkcolor=blue,urlcolor=blue]{hyperref}
\usepackage{tikz}
\usepackage{braket}
\usepackage{physics}
\usepackage{color}
\usepackage{soul}
\usepackage{mathtools}
\usepackage{float}
\usepackage{esvect}
\usepackage{stmaryrd} 
\usepackage{enumerate}
\usepackage{bbold}
\usepackage{ragged2e}
\usepackage{autobreak}

\makeatletter
\def\maketitle{
	\@author@finish
	\title@column\titleblock@produce
	\suppressfloats[t]}
\makeatother

\begin{document}

\begin{abstract}
\noindent 
We theoretically investigate the role of spatial dimension and driving frequency in a non-equilibrium phase transition of a driven-dissipative interacting bosonic system. In this setting,  spatial dimension is dictated by the shape of the external driving field. We consider both homogeneous driving configurations, which correspond to standard integer-dimensional systems, and fractal driving patterns, which give rise to a non-integer Hausdorff dimension for the spatial density. The onset of criticality is characterized by critical slowing down in the excited density dynamics as the system asymptotically approaches the steady state. By analyzing the system-size dependence of the asymptotic decay rate using numerical simulations of the full multi-mode dynamics, complemented by an analytical statistical mean-field treatment, we determine the lower critical dimension of the non-equilibrium phase transition. We show that this dimension can be non-integer and fractal in nature, and that it can be tuned continuously via the frequency detuning of the driving field.

\end{abstract}

\title{Tunable lower critical fractal dimension for a non-equilibrium phase transition}
\author{Mattheus Burkhard}
\affiliation{Universit\'{e} Paris Cit\'e, CNRS, Mat\'{e}riaux et Ph\'{e}nom\`{e}nes Quantiques, 75013 Paris, France}
\author{Luca Giacomelli}
\affiliation{Universit\'{e} Paris Cit\'e, CNRS, Mat\'{e}riaux et Ph\'{e}nom\`{e}nes Quantiques, 75013 Paris, France}
\author{ Cristiano Ciuti}
\affiliation{Universit\'{e} Paris Cit\'e, CNRS, Mat\'{e}riaux et Ph\'{e}nom\`{e}nes Quantiques, 75013 Paris, France}

\date{\today}

\maketitle

{\it Introduction ---}
A phase transition~\cite{ma2001modern} in a physical system manifests as a nonanalytic or discontinuous behavior of observable quantities that emerges in an appropriate thermodynamic limit. Spatial dimensionality plays a central role in determining critical behavior, with a particularly important concept being the lower critical dimension $d_L$: a phase transition that exists for dimensions $d \geq d_L$ is suppressed by fluctuations and ceases to occur for $d < d_L$. While equilibrium critical phenomena are well understood when restricted to integer spatial dimensions, the situation is far less settled for non-integer dimensions, especially in systems defined on fractal geometries~\cite{mandelbrot_fractal_1983,Fractal_Ising_2016,zhou_fractal_2021,myerson-jain_construction_2022}. In particular, studies of the equilibrium Ising model~\cite{Fractal_1980} have shown that critical behavior is not determined solely by the fractal (Hausdorff) dimension, but can also depend sensitively on additional geometrical characteristics of the underlying structure~\cite{Fractal_1980, Carmona_IsingSierpinski1998,Perreau_IsingFractal2017,Yi_QuantumIsingFractal2015}.

Further complications arise out of equilibrium, where phase transitions are substantially less explored and understood than their equilibrium counterparts. In non-equilibrium systems, a critical point is characterized by a critical slowing down of the dynamics as the steady state is approached in the thermodynamic limit. For open quantum systems~\cite{breuer_theory_2002}, this behavior is directly associated with the closing of the spectral gap of the Liouvillian governing the system dynamics~\cite{Minganti2018}.
Interacting optical systems~\cite{RMP2013} provide an especially powerful platform for investigating such transitions, owing to the precise control of external driving, nonlinearities, and dissipation. These ingredients compete to generate rich non-equilibrium dynamics and steady states~\cite{Kessler2012,carmichael_breakdown_2015,bartolo_exact_2016,Fink_PRX2017,biella_phase_2017,biondi_nonequilibrium_2017,Fitzpatrick2017,Vukics2019finitesizescalingof, FinkPRX2024,Clerk2025}. While for interacting bosonic systems such as photons the thermodynamic limit may also be achieved by taking the occupation number of a mode to be arbitrarily large~\cite{carmichael_breakdown_2015,Fink_PRX2017,Casteels2017,Rodriguez2017,Fink2017}, here we instead focus on the conventional thermodynamic limit of large spatial extent, which in optical settings corresponds to an infinite number of modes.

Recent studies of driven-dissipative interacting bosonic systems have revealed a first-order phase transition in two dimensions (2D) \cite{Foss-Feig2017,Vicentini2018,Li2022}, manifested by a discontinuous jump in the bosonic density. A theoretical analysis considered the regime in which the frequency detuning of the drive from the bosonic resonance is comparable to the dissipation rate and demonstrated that the system can display an emergent equilibrium behavior that maps onto the equilibrium Ising model \cite{Foss-Feig2017}. In this correspondence, the low and high density phases play the role of the paramagnetic and ferromagnetic phases, respectively. Since the Ising model exhibits a phase transition in 2D but not in 1D, an analogous dimensional dependence would be naturally expected for these driven dissipative systems. This expectation was subsequently experimentally confirmed on continuous platforms, where the detuning was set equal to the dissipation rate and the effective spatial dimension was engineered through a suitably shaped optical driving field, realizing one- and two-dimensional systems with diffusive boundary conditions \cite{Li2022}.

Despite these advances, fundamental questions concerning non-equilibrium critical phenomena in driven-dissipative systems remain open. In particular, the general role of spatial dimensionality is far from established, as existing studies have explored only a restricted region of parameter space. Crucially, previous investigations have focused on the specific regime in which the frequency detuning is close to the dissipation rate, leaving open the question of whether the observed dimensional dependence persists away from this fine-tuned condition. At the same time, continuous driven systems offer a unique degree of control over spatial structure, as the effective dimensionality can be engineered directly through the spatial profile of the driving field. This makes it possible to go beyond integer dimensions and to systematically investigate critical behavior in fractional, non-integer spatial dimensions, a regime that has so far remained unexplored.

\begin{figure*}[t!]
    \centering
    \includegraphics[width=0.95\linewidth]{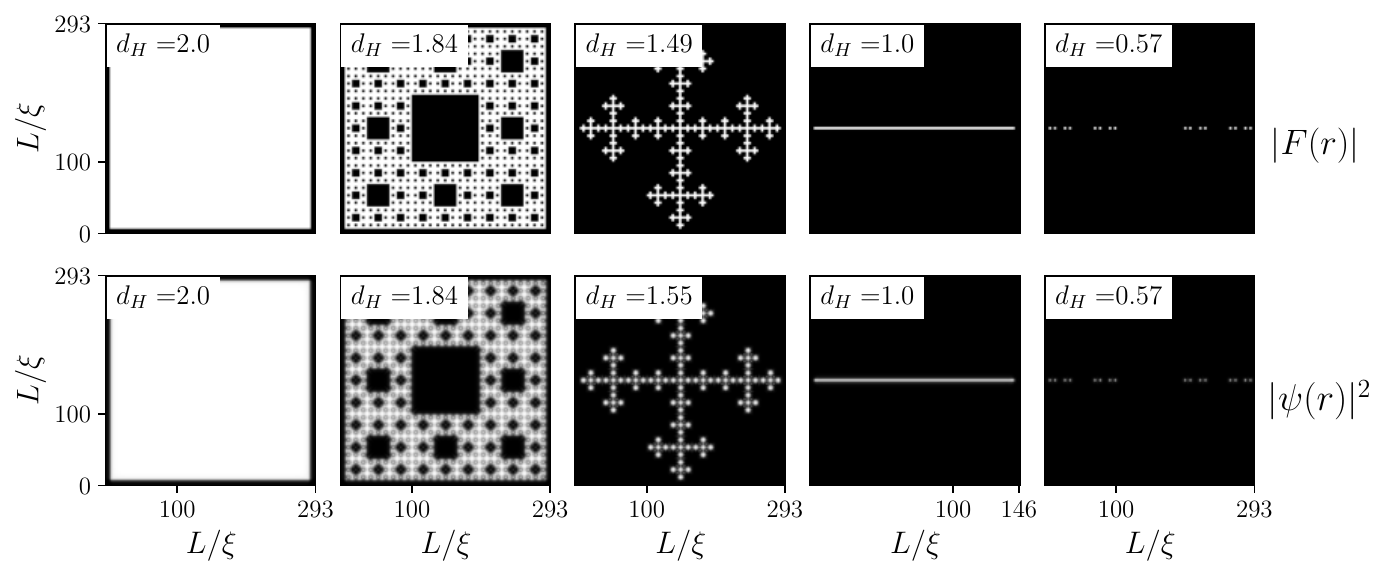}
    \caption{Driving intensity patterns (top panels) and grayscale maps (bottom panels) of the corresponding steady-state bosonic densities. Each column shows the spatial profile of a fractal pattern along with its associated Hausdorff box-counting dimension $d_H$. The first (fourth) column shows a 2D (1D) pattern with an integer dimension. The second (third) column displays results for a Sierpinski carpet (cross) after three iterations. The last column shows a system with dimension lower than one, a Cantor set with three iterations, corresponding to a system size with side length $L = 3^4l$, where $l$ is defined in the text. All figures obtained with the pump frequency detuning $\Delta=\gamma$ and with the driving amplitudes tuned to their respective critical points.}

    \label{fig:patterns-driving+densities}
\end{figure*}

In this Letter, we show that in a non-equilibrium setting the lower critical dimension is a tunable quantity and can take fractional values. We establish this by studying the phase transition of a driven dissipative, continuous, interacting two dimensional bosonic gas subjected to either homogeneous or fractal shaped driving fields, which impose different Hausdorff dimensions on the system. By analyzing the system size dependence of the long time relaxation rate toward the steady state, we demonstrate that the existence of a critical point is jointly controlled by the fractal dimension and the frequency detuning of the drive. These findings reveal that non equilibrium phase transitions can exhibit a level of richness and controllability that has no direct counterpart in equilibrium critical phenomena and remains largely unexplored.

{\it Theoretical framework ---} 
We consider a system of interacting bosons whose driven-dissipative dynamics can be described by a two-dimensional non-equilibrium Gross-Pitaevskii equation \cite{RMP2013, Li2022,Alexandre2014}, namely:
\begin{equation}
        i\frac{d\psi}{dt}  =\left[-\Delta-\frac{\hbar \nabla^2}{2m}+g\left|\psi\right|^2-i\frac{\gamma}{2}\right]\psi
        +F(\textbf{r}) \,.
        \label{eq:GPE}
\end{equation}
Here, $\psi(\textbf{r},t)$ denotes the mean value of the bosonic field operator, which depends on the spatial position $\mathbf{r}$ and the time $t$, and $n(\textbf{r},t)=|\psi(\textbf{r},t)|^2$ is the bosonic density. The equation is written in the so-called pump rotating frame, with $\Delta$ the detuning between the driving frequency and the frequency of the spatially uniform, lowest-frequency bosonic mode. Moreover, $g$ quantifies the nonlinear repulsive interaction, $\gamma$ is the natural dissipation rate of the system, and $F(\textbf{r})$ denotes the spatially dependent driving field. In the following, we express lengths in units of a characteristic length $\xi = \sqrt{\hbar/(2m\gamma)}$. In all simulations we use $g/(\gamma\xi^2)=0.0431$. Note that, within the non-equilibrium Gross-Pitaevskii framework,  changing its value is equivalent to a rescaling of the driving field and bosonic density, hence not affecting the presence or absence of a phase transition \cite{supp}.

In this class of systems, a phase transition is signaled by a sharp increase of the spatially averaged steady state density $\bar n_{ss}$ when the driving intensity $I=|F|^2$ reaches a critical value $I_c$ \cite{Foss-Feig2017,Vicentini2018,Li2022}. At the level of a mean-field description of a single-mode driven-dissipative bosonic system, no genuine phase transition is possible. Instead, for detunings $\Delta>\sqrt{3}/2 \gamma$, the mean-field equations display bistability between low and high-density steady state solutions \cite{RMP2013,Casteels2017,drummond_quantum_1980}. This bistable behavior does not correspond to a phase transition but represents a mean-field precursor of phase coexistence. Whether such coexistence survives and criticality emerges in the thermodynamic limit is decided by fluctuations associated with the full multi-mode dynamics. In particular, the criticality depends crucially on the spatial dimension of the system. Above a critical dimension, fluctuations are sufficiently suppressed and a discontinuous jump of the density persists in the thermodynamic limit, signaling a first-order phase transition from a dark low density state to a bright high density state.

The effective spatial dimensional can be controlled with the spatial shape of the driving field $F(r)$: pumping uniformly over a region of the plane results in a two-dimensional system with diffusive boundary conditions (first column of Fig.~\ref{fig:patterns-driving+densities}), while imposing a different geometry results in a lower dimension. For example, a pump profile that is narrow in one direction and elongated in the other creates an effectively one-dimensional system \cite{Li2022} (fourth column of Fig.~\ref{fig:patterns-driving+densities}). This can be extended to non-integer dimensions if the driving field has a fractal shape, as exemplified in the second and third columns of Fig.~\ref{fig:patterns-driving+densities}, where a Sierpinski carpet and a Sierpinski cross with a finite number of iterations are shown.

Note that the resulting bosonic density has in general a slightly distorted shape, as shown in the lower panels of Fig.~\ref{fig:patterns-driving+densities} (more patterns are shown in the Supplementary Material \cite{supp}). For example, in the second column, smaller holes in the Sierpinski carpet are partially filled and more affected than larger holes. This is due to interference and dispersive effects, as well as boson-boson interactions, that introduce intrinsic length scales in the system. Despite these distortions and the finite size, the resulting patterns possess their own fractal dimension, that can be quantified with the box-counting method \footnote{The box-counting method~\cite{Theiler_EstimatingDimension1990} quantifies the fractal dimension by overlaying a grid of $s^2$ square boxes, each of edge length $L/s$, where $L$ is the system size. After applying a thresholding algorithm~\cite{Otsu_Threshold1979} to extract the structure from a grayscale image, the number of boxes $A$ that intersect the pattern is counted. Assuming a scaling relation $A \sim s^{d_H}$, the fractal (Hausdorff) dimension $d_H$ is obtained as the slope of $\log A$ versus $\log s$ over a range of scales. More details are provided in the Supplementary Materials \cite{supp}.} and remain fractional and close to the one of the ideal infinite self similar pattern.
The box counting dimension $d_H$ of the driving patterns and the resulting densities are reported in Fig.~\ref{fig:patterns-driving+densities}.

While for integer-dimensional configuration the system size can be changed continuously, reaching the thermodynamic limit of a fractal system is less straightforward. Here, we consider the limit obtained by increasing the total system size while keeping the length of the smallest structural detail $l$ constant, as illustrated in Fig.~\ref{fig:ADR-vs-L}(a). This is important to set its ratio to the intrinsic physical lengths, which would otherwise drastically alter the box counting dimension of the resulting bosonic density. In all calculations we take $l = 10\mu m = 3.45\xi$ as an example such that $l>\xi$. This also implies that, at each step, the number of self-similar iterations is increased by one. This procedure results in a system size that increases by integer factors, as shown for the Sierpinski carpet in Fig.~\ref{fig:ADR-vs-L}(a), for which the linear size triples at each step. Although this exponential growth poses computational challenges, we were able to reach sizes that are much larger than the ones typically implemented in experimental realizations~\cite{Li2022}, and for which the critical behavior in 2D has already emerged.

{\it Phase transition markers ---}
In the investigation of phase transitions, several markers serve as indicators for identifying their presence or absence. In our case, one could use the maximal slope of the spatially averaged steady-state density as a function of the input intensity, that diverges when a first-order phase transition occurs~\cite{Li2022}. Another useful indicator is the asymptotic decay rate $\lambda$, also referred to as the Liouvillian gap in the Lindbladian framework \cite{Kessler2012,Minganti2018}, that is the decay rate with which the system relaxes exponentially towards its steady state at long times. At the critical point of a driven-dissipative phase transition, critical slowing down occurs, that is the system requires an infinite amount of time to reach its steady state, and $\lambda$ tends to zero in the thermodynamic limit~\cite{Minganti2018, Casteels2017}.
Conversely, if in this infinite-size limit the decay rate remains finite for all values of the driving intensity, no phase transition occurs.

The asymptotic decay rate $\lambda$ can be obtained from the time-dependence of the spatially averaged density $\bar{n}(t)=\int \mathrm{d}^2x\ n(t,x)$, that at long times evolves as ${\bar{n}(t) - \bar{n}_{ss} \propto e^{-\lambda t}}$, where $\bar{n}_{ss}$ is the steady-state value. We compute this value by numerically evolving the dynamical equation~(\ref{eq:GPE}) and performing a Fourier analysis of $\bar n(t)$~\cite{supp}.  This procedure is repeated for different driving intensities $I$ to find the minimal asymptotic decay rate $\lambda_{\min}=\min_I \lambda(I)$. To probe the presence of a critical behaviour, we compare the values of $\lambda_{\min}$ obtained for different system sizes: a critical point is present if this quantity tends to zero for increasing system sizes, while it is absent if if saturates to a finite value.

\begin{figure}[t!]
    \centering
    \includegraphics[width=\linewidth]{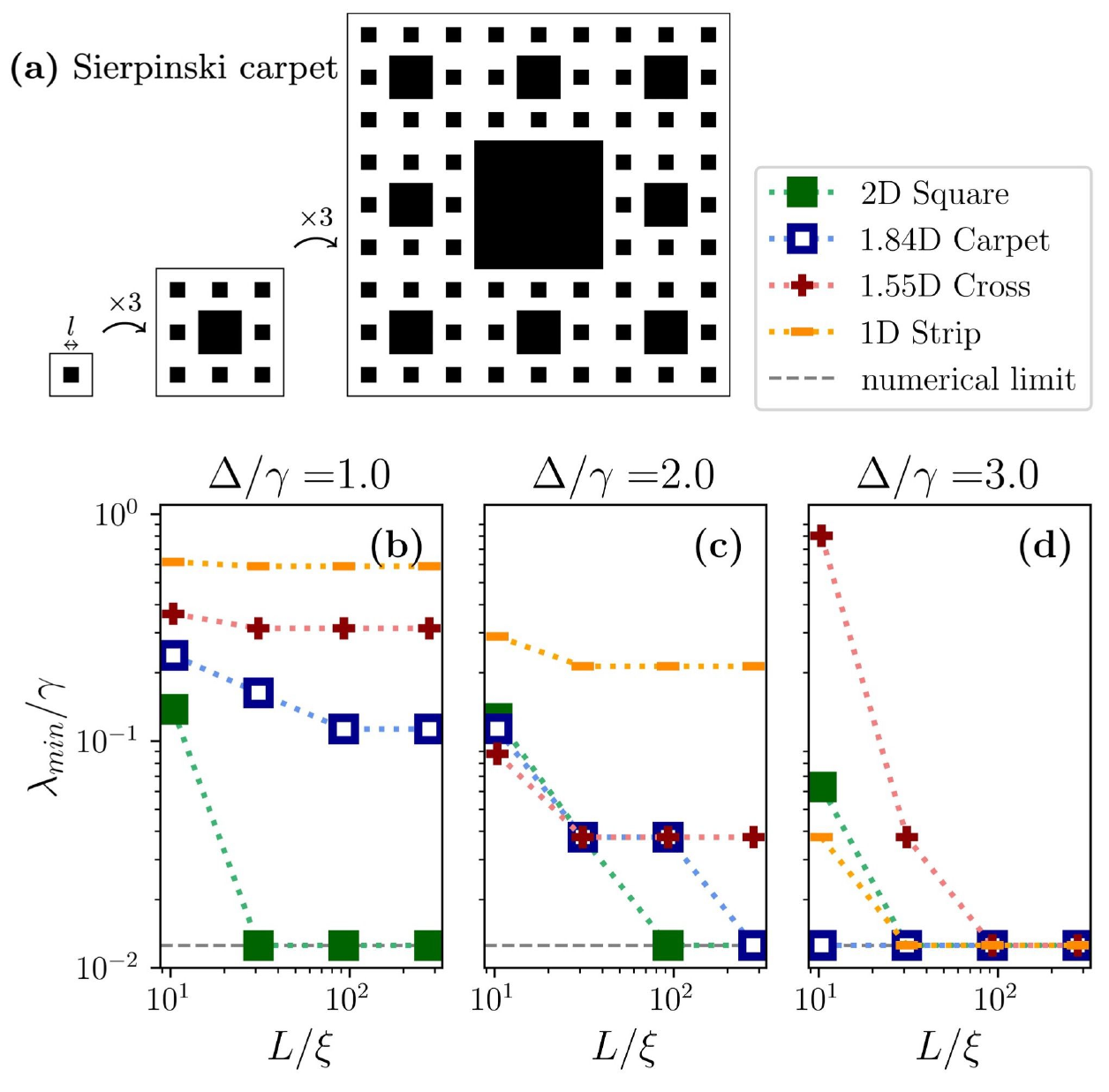} 
   \caption{(a) Illustration of the approach to the thermodynamic limit by expanding the size of a fractal pattern. At each step, the total linear size $L$ is multiplied by three, while the smallest resolved feature $l$ remains constant. (b–d): Minimal asymptotic decay rate $\lambda_{\min}$ as a function of the linear system size $L$ for detuning values $\Delta = 1.0\gamma$, $\Delta = 2.0\gamma$, and $\Delta = 3.0\gamma$. Different markers correspond to different spatial geometries indicated in the legend. The grey dashed line indicates the numerical limit $1/t_f$, which corresponds to the frequency resolution of the simulations set by the finite evolution time $t_f$.
    }
    \label{fig:ADR-vs-L}
\end{figure}

{\it Results at fixed detuning ---}
We begin by examining the case $\Delta = \gamma$, for which in previous investigations the first-order phase transition was found to occur in 2D, and to be absent in 1D~\cite{Foss-Feig2017,Vicentini2018,Li2022}. This is confirmed by our calculations of $\lambda_{\min}$, whose dependence on the system size is shown in Fig.~\ref{fig:ADR-vs-L}(b), with green squares for a 2D system and orange bars for a 1D system. In fact, for the 2D case and in the limit of large systems, $\lambda_{\min}$ vanishes {(with a precision set by the length $t_f$ of the time evolution).
This critical slowing down is instead absent in the one-dimensional system, as $\lambda_{\min}$ saturates to a finite value.

We extended this investigation to fractal dimensions. For example, for the Sierpinski carpet (second column of Fig.~\ref{fig:patterns-driving+densities}) the asymptotic decay rate has a minimum around the expected mean-field critical driving \cite{supp}, but $\lambda_{\min}$ converges to a finite value for increasing system size, as shown in Fig.~\ref{fig:ADR-vs-L}(b) with empty blue square markers. The system therefore exhibits slowing down, but it is not critical, so that, at least for this value of the detuning, the Sierpinski carpet of dimension $d_H\simeq 1.84$ is below the critical dimension. This reflects also on the steady-state average density, that does not display a discontinuous jump as a function of $I$, and has instead a smooth crossover from the low density to the high density phase \cite{supp}. We obtain a similar behavior for the Sierpinski cross (third column of Fig.~\ref{fig:patterns-driving+densities}), which has $d_H \simeq 1.5$ and whose convergence of $\lambda_{\min}$ to a finite value is shown with red cross markers in Fig.~\ref{fig:ADR-vs-L}(b). It is interesting to comment that the shape the longest-lived mode associated to $\lambda_{\min}$ in a fractal system is non-trivial (see Figure in~\cite{supp}) and will be the object of further studies.

\begin{figure}[t!]
    \centering
    \includegraphics[width=\linewidth]{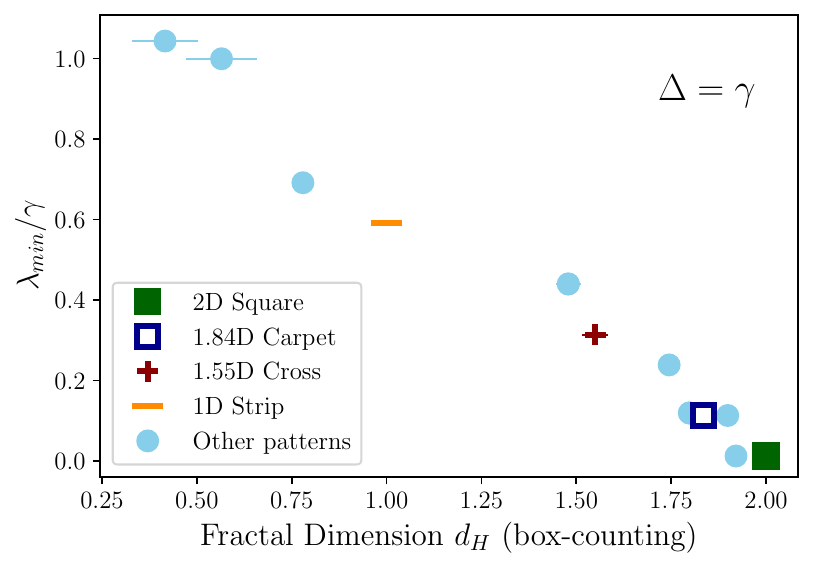} 
    \caption{Minimal asymptotic decay rate $\lambda_{\min}$ versus Hausdorff dimension for a fixed detuning $\Delta=\gamma$. The patterns for the light blue circular markers are shown in \cite{supp}. Horizontal error bars show the uncertainty of the box-counting method, which is more significant only for patterns of fractal dimension smaller than one (Cantor sets).}
    \label{fig:ADR-vs-dH}
\end{figure}

Besides the fact that it vanishes or not, the value of $\lambda_{\text{min}}$ also quantifies how far the system is from criticality. We found it to depend strongly on the system's dimensionality, as shown in Fig.~\ref{fig:ADR-vs-dH}, where we plot $\lambda_{\text{min}}$ as a function of the Hausdorff dimension $d_H$, for a variety of fractal geometries. These values are the ones for the largest computalionally feasible sizes (ranging from $L=81l$ to $L=216l$, depending on the geometry), and are representative of the thermodynamic limit $\lim_{L \to \infty} \lambda_{\text{min}}$. In fact, the difference between $\lambda_{\min}$ for the two largest considered sizes gives us a vertical error bar that is always smaller than the marker size. Remarkably, $\lambda_{\min}$ exhibits a clear monotonous dependence on the Hausdorff dimension $d_H$. The slowing down continuously approaches a critical one while the dimension is increased, that in this case $\Delta=\gamma$ is only reached when the dimension is close to 2, i.e. the lower critical dimension for the first-order phase transition is $d_L(\Delta = \gamma) \approx 2$. This extends previous results~\cite{Foss-Feig2017,Vicentini2018,Li2022} to fractional dimensions, but, as we show in the following, the lower critical dimension surprisingly decreases significantly from 2 as $\Delta$ increases.

\begin{figure}[t!]
    \centering
    \includegraphics[width=\linewidth]{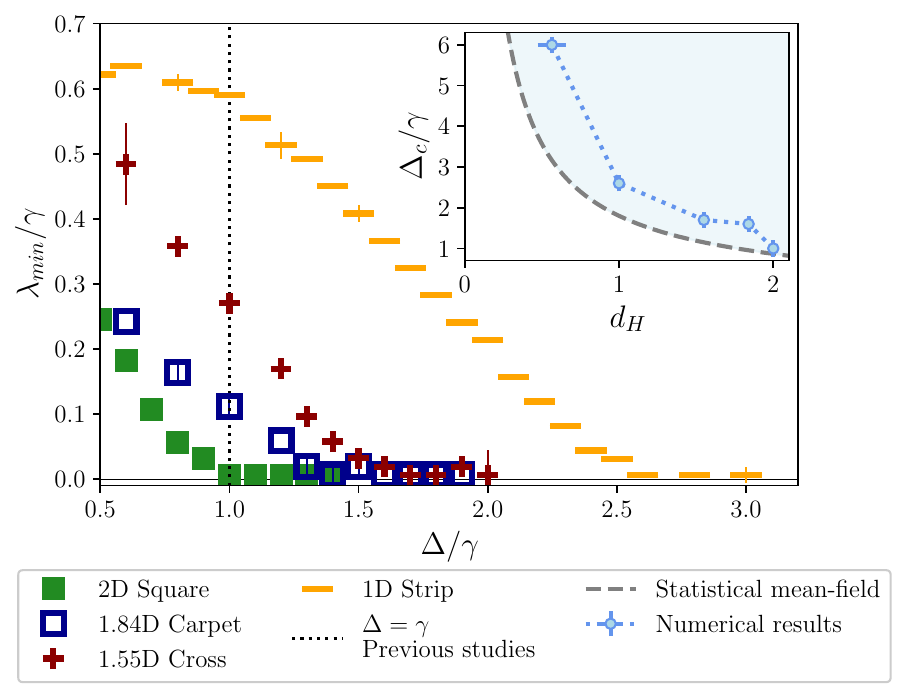}
     \caption{Minimal asymptotic decay rate as a function of detuning $\Delta$ for different spatial dimensions. All data points are obtained for a system size $L = 81l$. The vertical error bars represent the difference in $\lambda_{\min}$ between this size and $L = 27l$, providing an estimate of the finite-size effects and our proximity to the thermodynamic limit.
Inset: Critical detuning $\Delta_c$ as a function of the Hausdorff dimension $d_H$. Error bars correspond to the numerical resolution. The gray dashed line shows the critical detuning predicted by the statistical mean-field approach (see main text), above which the phase transition is possible (shaded area). }
    \label{fig:ADR-vs-Delta}
\end{figure}

{\it Changing the lower critical dimension ---}
As shown in Figs.~\ref{fig:ADR-vs-L}(c)-(d) and in Fig.~\ref{fig:ADR-vs-Delta}, increasing the driving detuning $\Delta$ leads to a systematic decrease of $\lambda_{\min}$, and eventually to its vanishing, for all geometries. This implies that the system is critical and exhibits a first-order phase transition even for dimensions  $d_H < 2$, provided that the ratio $\Delta/\gamma$ is large enough. For a fixed dimension $d_H$, we define the critical detuning $\Delta_c$ as the value above which $\lambda_{\min}$ vanishes in the thermodynamical limit.
For the 2D system (green squares), $\Delta_c \simeq \gamma$, close to the previously studied case. $\Delta_c$ is instead larger for lower-dimensional systems. This can be seen in Fig.~\ref{fig:ADR-vs-L}(c), where, for large enough systems, $\lambda_{\min}$ vanishes not only for the 2D square, but also for the Sierpinski carpet, which for $\Delta=\gamma$ is not critical. Correspondingly, the smooth crossover in the average density for $\Delta=\gamma$, becomes a sharp discontinuity \cite{supp}. For $\Delta=2\gamma$ the lower-dimensional Sierpinski cross and the one-dimensional system are still non-critical, but both become for a higher $\Delta=3\gamma$, as shown in Fig.~\ref{fig:ADR-vs-L}(d). These results show that it is possible to tune the lower critical dimension, or, in other words, to make the phase transition occur at any fixed dimension by increasing $\Delta$. The inset of Fig.~\ref{fig:ADR-vs-Delta}, shows that the lower critical dimension decreases continuously as the detuning is increased. Interestingly, also the 1D case and fractals with dimension lower than one (e.g., Cantor sets) can be made critical by increasing $\Delta/\gamma$ enough. In fact, in the limit $d_H \to 0^+$, we can expect $\Delta_c \to + \infty$, since for a zero-dimensional (one-mode) nonlinear bosonic resonator it is known that the asymptotic decay rate vanishes only in the limit of infinite detuning~\cite{Rodriguez2017,Casteels2017-power_laws}. 

To complement our numerical results, we gain insight into the existence of a critical detuning by introducing an analytical statistical mean-field theory that incorporates the effect of kinetic energy contributions in the non-equilibrium Gross-Pitaevskii equation \eqref{eq:GPE}. For a fractal driving pattern, the system contains multiple interfaces between pumped and non-pumped regions, such that the kinetic energy contribution associated with these gradients becomes important. We therefore construct the statistical mean-field approximation by incorporating the kinetic contribution through  the energy cost of a single interface, multiplied by the average number of interfaces within a region of size comparable to the smallest feature of the fractal pattern. This number can be related to the Hausdorff dimension through the average connectivity $\tilde z=2d_H$, while an analytical estimate of the kinetic contribution can be obtained from a solution of \eqref{eq:GPE} with $g=0$. This derivation, detailed in the End Matter, yields an equation for the steady-state density with an effective, dimensionality-dependent detuning, which displays bistability when
\begin{equation}
\Delta_{eff}=\Delta-\frac{4-2d_H}{4}\sqrt{\Delta^2+\frac{\gamma^2}{4}}>\frac{\sqrt{3}}{2}\gamma.
\end{equation}
This provides an estimate of the critical detuning, shown as the dashed gray line in the inset of Fig.~\ref{fig:ADR-vs-Delta}. Since bistability is a necessary condition for the emergence of critical behavior, this prediction constitutes a lower bound for the critical detuning. It captures well the overall trend of the numerical data obtained from brute-force simulations of the full multimode dynamics, that lie slightly above this lower bound. This shows that, upon decreasing the dimensionality, the phase transition is suppressed at sufficiently low detuning, leading to a lower critical dimension that can be fractal.

{\it Conclusions ---}
In conclusion, we have demonstrated that in driven-dissipative systems of interacting bosons, the lower critical dimension can be continuously tuned with the frequency detuning of the driving field and can be fractional. This establishes a novel paradigm in the study of non-equilibrium phase transitions, where geometry and external drive jointly control critical behavior.

Our results were obtained by solving the multi-mode mean-field dynamics and analyzing the emergence of critical slowing down as the size of the driving patterns increases, and are supported by an analytical statistical mean-field theory that captures the dependence of the critical detuning on the Hausdorff dimension. An open and important question concerns the fate of the lower critical dimension in the presence of strong quantum boson-boson interactions, beyond the many-mode mean-field regime, particularly in the hard-core boson limit. Three scenarios could take place: (i) the lower critical dimension is the same in the strongly correlated regime; (ii) the lower critical dimension could evolve continuously with interaction strength, or (iii) a discontinuous shift could occur, signaling the emergence of a quantum phase transition with new phases in the strongly quantum correlated regime. Exploring this limit remains a key direction for future research. Another promising avenue is to investigate how fractional dimension and driven-dissipative conditions influence other classes of non-equilibrium phase transitions. This could yield deeper insight into the fundamental role of spatial geometry and dissipation in shaping critical phenomena in classical and quantum open systems.

\acknowledgements{We acknowledge support from the French ANR project FracTrans (grant ANR-24-CE30-6983) and  a grant (Polaritonic) from the French Government managed by the
ANR under the France 2030 programme with the reference ANR-24-RRII-0001. This work was also funded by the French ministry of research through a CDSN
grant of ENS Paris-Saclay.}

\bibliography{fractal.bib}

\newpage
\clearpage

\onecolumngrid


\begin{center}
{\textbf{End Matter}}
\end{center}

\twocolumngrid

{\it Statistical mean field approach ---}  
The dissipative phase transition is related to the bistable behavior of the bosonic system, which is predicted within a single-mode mean-field approximation. In this framework, the Gross-Pitaevskii equation~(\ref{eq:GPE}) in the steady state and for a homogeneous 2D driving reduces to
\begin{equation}
    0=\left[-\Delta+g\left|\psi\right|^2-i\frac{\gamma}{2}\right]\psi
        +F \,.
\end{equation}
By isolating the driving term on the left-hand side and taking the modulus squared, one obtains, with $n=|\psi|^2$,
\begin{equation}
        I=|F|^2=\left[(gn-\Delta)^2+\frac{\gamma^2}{4}\right]n\,.
    \label{eq:bistability}
\end{equation}
This nonlinear relation determines the possible steady-state solutions for the density $n$ at a given driving intensity $I$ and can be interpreted as an equation of state~\cite{RMP2013}. For $\Delta>\sqrt{3}/2~\gamma$, this equation predicts a bistable behavior, namely the coexistence of two stable density solutions over a finite range of driving intensities~\cite{drummond_quantum_1980}. While this bistability does not correspond to a phase transition in a single-mode system, it constitutes a necessary condition for the appearance of the dissipative phase transition investigated here. In the homogeneous two-dimensional case, this mean-field analysis yields a critical detuning $\Delta_c^{MF}(d_H=2)=\sqrt{3}/2~\gamma$.

To incorporate the role of dimensionality and extend this type of analysis to fractal systems, we introduce what we refer to as a statistical mean-field approach. We start by writing the functional
\begin{align*}
\frac{1}{\hbar} \mathcal{F}[\psi]
=\int \! dx\,dy \,
\bigg[
& - \Delta |\psi|^2
+ \frac{\hbar}{2m} \, |\nabla \psi|^2 \\
& + \frac{g}{2} |\psi|^4
- i \frac{\gamma}{2} |\psi|^2
+ F \left(\psi^*+\psi\right)
\bigg]\;,
\end{align*}
whose stationarization yields the Gross-Pitaevskii equation
\begin{align*}
\frac{1}{\hbar}\,\frac{\delta \mathcal{F}}{\delta \psi^*} &= 0\\
&= - \Delta \psi -\frac{\hbar}{2m} \nabla^2 \psi
- i \frac{\gamma}{2} \psi
 + g |\psi|^2 \psi
+ F ~.
\end{align*}
Our goal is to approximate the full functional $\mathcal{F}[\psi]$ with a coarse-graining that explicitly accounts for the spatial structure of the driving pattern. Concretely, we partition space into square patches of size $l\times l$, where $l$ is the smallest characteristic length scale of the fractal pattern, and write the functional as
\begin{align*}
\frac{1}{\hbar} \mathcal{F}[\psi]
=
\sum_p\int_0^l dx_p \int_0^l dy_p
\bigg[ &
- \Delta |\psi|^2 +
\frac{\hbar}{2m} \, |\nabla\psi|^2 \\
& + \frac{g}{2} |\psi|^4 
- i \frac{\gamma}{2} |\psi|^2
+ F\psi^*
\bigg]~,
\end{align*}
where the sum runs over the patches and the subscript $p$ labels quantities within each patch.

We assume the field to vanish in the non-driven patches ($F(x_p,y_p)=0$) and to be approximately constant, $\psi(x_p,y_p)\simeq \phi$, in the driven ones ($F(x_p,y_p)=F$). The kinetic term requires special care, as it contributes only at the interfaces between driven and undriven regions, where the field varies spatially. An analytical expression for this kinetic contribution can be obtained by computing the energy cost associated with such an interface neglecting $g \vert \Psi \vert^2$. This is a good approximation because the inclusion of this term only slightly distorts the linear solution for the density near the pump edge. This amounts to solving
\[
\frac{\hbar}{2m}\partial_x^2\psi(x)=-\left(\Delta+i\frac{\gamma}{2}\right)\psi(x) +F(x)
\]
for a step-like driving field $F(x)=\theta(-x)F$. The solution can be expressed as the sum of the constant bulk solution $\phi$ in the driven region and a plane-wave contribution with amplitude $\frac{\phi}{2}$ and momentum
$$
k^2=\frac{2m}{\hbar}\left(\Delta+i\frac{\gamma}{2}\right).
$$
By approximating all interfaces as behaving identically, the kinetic gradient term can thus be replaced by $|\nabla\psi|^2\simeq |k|^2|\frac{\phi}{2}|^2$. Counting the number of edges associated with each patch $p$ then yields a contribution proportional to $(z-z_p)$, where $z=4$ is the coordination number of the underlying two-dimensional lattice and $z_p$ is the local coordination number, defined as the number of neighboring driven patches. The functional can therefore be approximated by the following sum over driven patches:
\begin{align*}
    \frac{1}{\hbar} \mathcal{F}[\psi]
\approx
\sum_p l^2
\bigg[&
- \Delta |\phi|^2
+ (z-z_p)\frac{\hbar}{8m} \, |k|^2|\phi|^2\\
& + \frac{g}{2} |\phi|^4
- i \frac{\gamma}{2} |\phi|^2
+ F \phi^*
\bigg]~.
\end{align*}

We can now introduce a rescaled and spatially averaged functional by defining the average coordination number $\tilde{z}$ associated with the specific driving pattern:
\[
f(\phi)
=
- \Delta |\phi|^2
+(z-\tilde{z})\frac{\hbar}{8m} |k|^2 |\phi|^2
- i \frac{\gamma}{2} |\phi|^2
+ \frac{g}{2} |\phi|^4
+ F \phi^*~.
\]
Extremizing this with respect to $\phi^*$ yields the \textit{statistical} mean-field steady-state equation
\begin{equation}
    \dot\phi=0=\left[-\Delta+(z-\tilde{z})\,\eta(\Delta)+g|\phi|^2-i\frac{\gamma}{2}\right]\phi+F,
\end{equation}
where we introduced the interface energy scale
\begin{equation}
\eta(\Delta)=\frac{\hbar}{8m} |k|^2=\frac{1}{4}\sqrt{\Delta^2+\gamma^2/4}~.
\end{equation}
From this expression, we can derive the corresponding equation of state
\begin{equation}
    I=\left[ \left(g|\phi|^2 -\Delta+(z-\tilde{z})\,\eta\right)^2+\frac{\gamma^2}{4} \right]|\phi|^2.
    \label{eq:bistability bis}
\end{equation}
This formulation extends the usual single-mode description of a uniform system by incorporating an additional term $(z-\tilde{z})\,\eta$ that accounts for the energetic cost of interfaces. We can further relate the average connectivity $\tilde{z}$ to the fractal dimension through the relation $\tilde{z}=2d_H$, which holds for integer-dimensional systems.

As in Eq.~\eqref{eq:bistability}, this equation of state exhibits bistability when
\begin{equation}
    {\Delta_{eff}=\Delta-(4-2d_H)\,\eta(\Delta)}>\frac{\sqrt{3}}{2}\gamma~. 
\end{equation}
Since bistability is a necessary condition for the existence of a phase transition, this relation provides a dimensionality-dependent estimate of the critical detuning below which the phase transition cannot occur,
\begin{equation}\label{eq:mf-crit-detuning}
\begin{split}
    \Delta_c^{MF}(d_H)=&
     \Delta_c^{MF}(2)+(z-\tilde{z}) \, \eta\big(\Delta_c^{MF}(d_H)\big)\\
     =&\frac{\sqrt{3}}{2}\gamma+(4-2d_H)\,\eta\big(\Delta_c^{MF}(d_H)\big)\\
     =&\frac{\sqrt{3}}{2}\gamma+\frac{(4-2d_H)}{4}\sqrt{\big(\Delta_c^{MF}(d_H)\big)^2+\frac{\gamma^2}{4}}~.
\end{split}
\end{equation}
This dependence is shown as a dashed line in the inset of Fig.~\ref{fig:ADR-vs-Delta}. The resulting curve follows the same qualitative trend as the numerical data while remaining below them, as expected for a lower bound. Moreover, Eq.~\eqref{eq:mf-crit-detuning} predicts that the critical detuning diverges as the dimensionality approaches zero, in agreement with the behavior found in single-mode Kerr resonators \cite{Rodriguez2017}.

\newpage
\clearpage

\title{{\bf Supplementary Material for the article:}\\ ``Tunable lower critical fractal dimension for a non-equilibrium phase transition"}
\setcounter{page}{1}
\setcounter{equation}{0}
\setcounter{figure}{0}
\renewcommand{\theequation}{S\arabic{equation}}
\renewcommand{\thefigure}{S\arabic{figure}}
\date{\today}
\maketitle
\onecolumngrid

This Supplementary Material provides further theoretical details and additional numerical results that support and extend the findings presented in the main text.

\section{Numerical method}

We numerically simulate the driven--dissipative Gross--Pitaevskii equation [Eq.~(\ref{eq:GPE})] on a two-dimensional spatial grid. The system is initialized in the vacuum state, $\psi(\mathbf{r},t=0)=0$ for all $\mathbf{r}$. At time $t=0$, the coherent pump is switched on with a fixed driving amplitude $F$, corresponding to a driving intensity $I = F^2$. The equation is then time evolved up to a final time $t_f$, chosen sufficiently large for the dynamics to converge to the steady state. Time integration is performed using a second-order Runge--Kutta scheme with a fixed time step selected to ensure numerical stability.

To characterize the long-time dynamics, we compute the spatially averaged density $\bar{n}(t)=\int \mathrm{d}^2x\, n(\mathbf{r},t)$ and apply a Fourier transform to its time evolution. From the resulting spectrum, we extract the asymptotic decay rate $\lambda$ by measuring the full width at half maximum (FWHM) of the central peak, which provides the slowest relaxation rate of the system.

We have verified the robustness of our results by including quantum fluctuations within the Truncated Wigner Approximation, as shown in Fig.~\ref{fig:TWA}. The driven-dissipative Gross--Pitaevskii equation is found to capture the  behavior remarkably well, as the spatial inhomogeneities induced by the driving pattern effectively act as \textit{spatial fluctuations}~\cite{Li2022}.
\begin{figure}
    \centering
    \includegraphics[width=\linewidth]{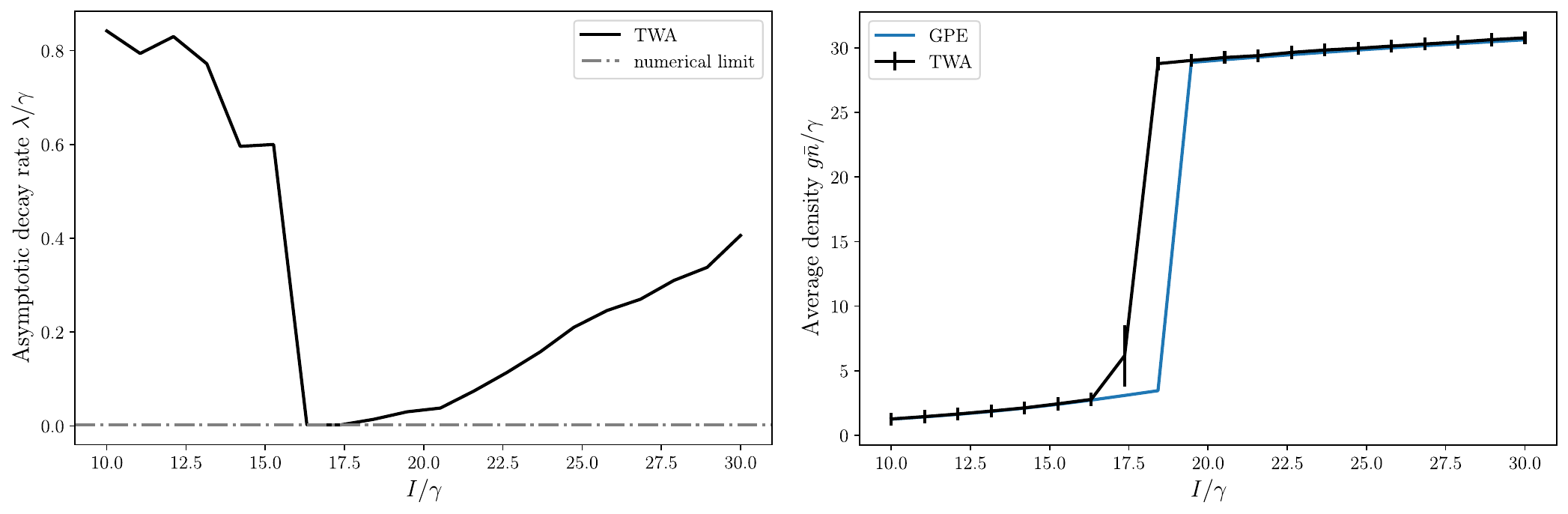}
    \caption{First-order phase transition observed within the Truncated Wigner Approximation. Here, we consider the same 1D system as in the main text, with a longitudinal size $L=34.5\xi$ and $\Delta=3\gamma$. The observables are averaged over 100 trajectories.}
    \label{fig:TWA}
\end{figure}

\section{The effect of detuning for a Sierpinski carpet}

In the main text, we characterize critical behavior in terms of the minimal asymptotic decay rate. Here, we further examine how the asymptotic decay rate depends on the driving intensity for different system sizes. In Fig.~\ref{fig:SC-PT markers}(a)--(b), these quantities are shown for the Sierpinski carpet (second column in Fig.~\ref{fig:patterns-driving+densities}) with $\Delta=\gamma$, a regime where no phase transition occurs. The minimal asymptotic decay rate exhibits a clear minimum around a critical driving intensity $I\approx2.2\gamma^2$, yet this minimum saturates to a finite value.

\begin{figure}[t]
    \centering
    \includegraphics[width=\columnwidth]{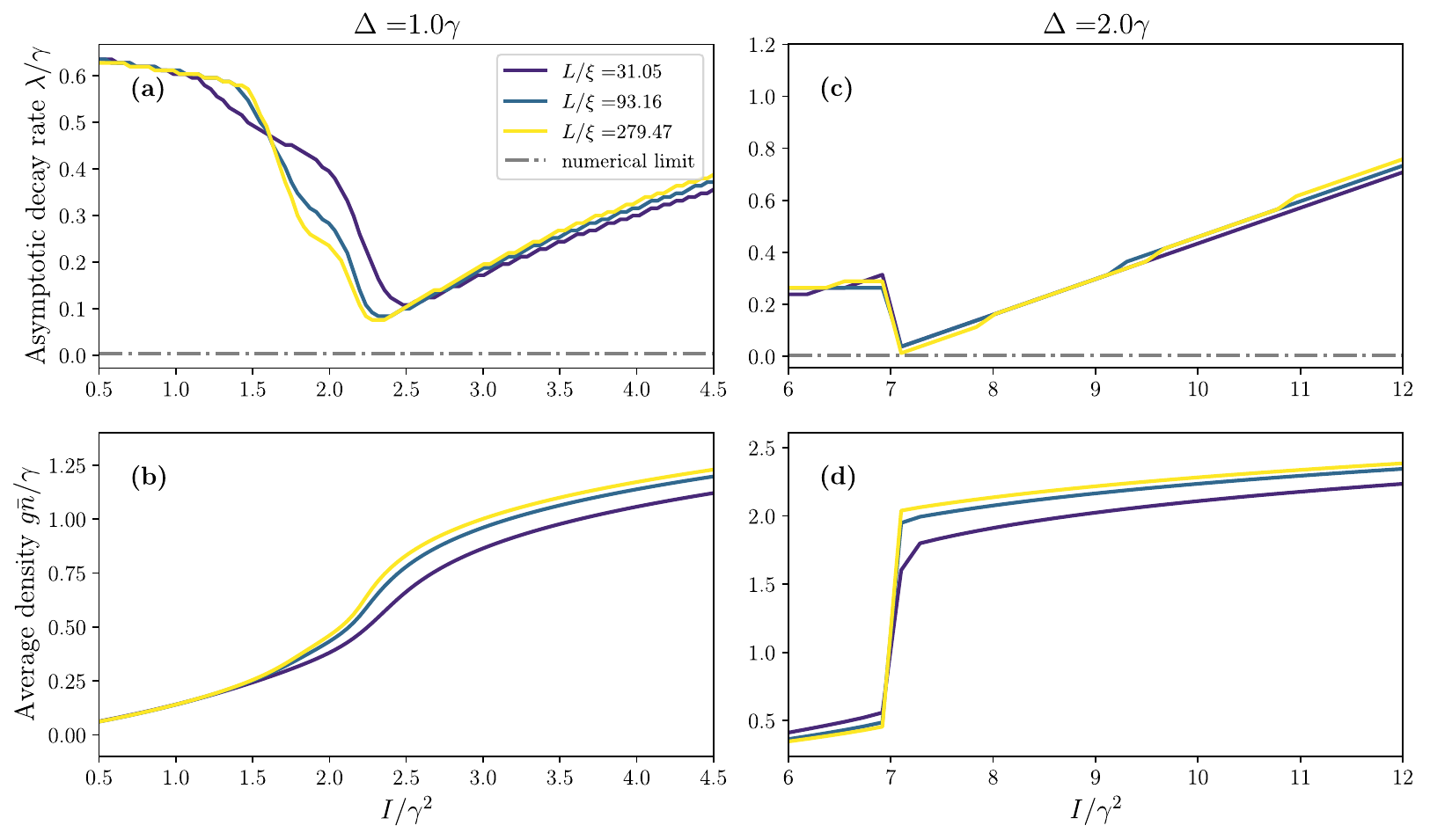}
    \caption{Response of the system to a Sierpinski carpet drive with fractal dimension $d_H = 1.84$. 
(a) Asymptotic decay rate as a function of input intensity for detuning $\Delta = \gamma$, showing no phase transition. 
(b) Corresponding bosonic density as a function of input intensity, with a smooth increase. 
(c) For a larger detuning $\Delta = 2.0\gamma$, the minimal asymptotic decay rate $\lambda_{\min}$ vanishes at the critical point $I_c \approx 7.1\gamma^2$, indicating a first-order phase transition. 
(d) At the same critical point, the average bosonic density exhibits a discontinuous jump.}

    \label{fig:SC-PT markers}
\end{figure}

In Fig.~\ref{fig:SC-PT markers}(c)--(d), we plot the same quantities for a larger detuning, $\Delta=2\gamma$. In this case, at a higher driving intensity $I_c\approx7.1\gamma^2$, the minimal asymptotic decay rate $\lambda_{\min}$ collapses to numerical precision, and the density exhibits an abrupt switch from the lower to the upper branch of the bistability. These features indicate the occurrence of a first-order phase transition.

\section{Fractal dimension : nominal dimension vs box-counting dimension}

In this section, we discuss different notions of fractal dimension and clarify the distinction between nominal and box-counting definitions. The nominal mathematical fractal dimension \( d \) of a Sierpinski-carpet-like object can be determined from its self-similar scaling properties. If the pattern is constructed by dividing a square into \( s \times s \) smaller squares and removing a given number of them (the “holes”), then at each iteration the remaining area scales as
\[
A = s^2 - N_{\text{holes}}.
\]
This area corresponds to the white surface shown in Fig.~\ref{fig:dimension schema}. The nominal fractal dimension is then obtained from the ratio of logarithms,
\[
d^{th}_H = \frac{\log(A)}{\log(s)} = \frac{\log(s^2 - N_{\text{holes}})}{\log(s)}.
\]
This expression characterizes how the number of occupied sites, or equivalently the total area, scales with the rescaling factor \( s \) for a \textit{theoretical} infinitely self-similar fractal. The procedure is illustrated in Fig.~\ref{fig:dimension schema} for the specific case of the Sierpinski carpet.

\begin{figure}[h]
\centering
\begin{tikzpicture}[scale=0.95]
\begin{scope}
    \draw (0,0) rectangle (3,3);

    \node at (1.5,-0.5) {simple square} ;
\end{scope}

\begin{scope}[shift={(5,0)}]
    \draw (0,0) rectangle (3,3);
    \draw [fill=black] (1,1) rectangle (2,2) ;
    \draw [thin] (0,1) -- (3,1);
    \draw [thin] (0,2) -- (3,2);
    \draw [thin] (1,0) -- (1,3);
    \draw [thin] (2,0) -- (2,3);
    \node at (1.5,-0.5) {1st iteration} ;
\end{scope}

\begin{scope}[shift={(10,0)}]
    \draw (0,0) rectangle (3,3);

    \draw [fill=black] (1/3,1/3) rectangle (2/3,2/3) ;
    \draw [fill=black] (1+1/3,1/3) rectangle (1+2/3,2/3) ;
    \draw [fill=black] (2+1/3,1/3) rectangle (2+2/3,2/3) ;
    \draw [fill=black] (1/3,1+1/3) rectangle (2/3,1+2/3) ;
    \draw [fill=black] (1/3,2+1/3) rectangle (2/3,2+2/3) ;
    \draw [fill=black] (2+1/3,2+1/3) rectangle (2+2/3,2+2/3) ;
    \draw [fill=black] (2+1/3,1+1/3) rectangle (2+2/3,1+2/3) ;
    \draw [fill=black] (1+1/3,2+1/3) rectangle (1+2/3,2+2/3) ;
    
    \draw [fill=black] (1,1) rectangle (2,2) ;
    \draw [thin] (0,1) -- (3,1);
    \draw [thin] (0,2) -- (3,2);
    \draw [thin] (1,0) -- (1,3);
    \draw [thin] (2,0) -- (2,3);
    
    \draw [ultra thin] (0,1/3) -- (3,1/3);
    \draw [ultra thin] (0,2/3) -- (3,2/3);
    \draw [ultra thin] (1/3,0) -- (1/3,3);
    \draw [ultra thin] (2/3,0) -- (2/3,3);

    \draw [ultra thin] (0,1+1/3) -- (3,1+1/3);
    \draw [ultra thin] (0,1+2/3) -- (3,1+2/3);
    \draw [ultra thin] (1+1/3,0) -- (1+1/3,3);
    \draw [ultra thin] (1+2/3,0) -- (1+2/3,3);

    \draw [ultra thin] (0,2+1/3) -- (3,2+1/3);
    \draw [ultra thin] (0,2+2/3) -- (3,2+2/3);
    \draw [ultra thin] (2+1/3,0) -- (2+1/3,3);
    \draw [ultra thin] (2+2/3,0) -- (2+2/3,3);
    
    \node at (1.5,-0.5) {2nd iteration} ;
\end{scope}
\end{tikzpicture}

\caption{Constructing the Sierpinski Carpet. We divide an initial square in a grid of $3\times3$ squares. We remove the central one. Then we can repeat this for the remaining eight squares. The area after one iteration is $A=s^2-N_{\text{holes}}=3^2-1=8$. So the mathematical theoretical dimension is $d^{th}_H=\log(8)/\log(3)=1.89$.}

\label{fig:dimension schema}
\end{figure}

In a purely mathematical context, two different fractals can share exactly the same fractal dimension. This occurs, for example, when one chooses the same subdivision factor, $s$ and the same number of removed sites $N_{\text{holes}}$, but removes them in a different geometric arrangement at each iteration. One could follow the same construction procedure as in Fig.~\ref{fig:dimension schema}, but instead remove, say, the upper-left square rather than the central one; the resulting object would have a drastically different visual appearance (see Fig.~\ref{fig:disconnected_pattern} (left)), yet its nominal dimension $d^{th}_H$ would remain identical because the scaling rule encoded in $\log(A)/\log(s)$ is unchanged. 

However, it is crucial to understand that in our physical system the final spatial profile of the bosonic density does not necessarily inherit this mathematical invariance: distinct driving geometries can lead to patterns that exhibit different effective fractal dimensions. This arises because the pattern is subject to physical distortions—such as nonlinear interactions, finite-resolution effects, and the intrinsic dynamics of the driven–dissipative medium—that modify how the nominal fractal mask is mapped onto the actual stationary density distribution. As a result, two drivings with mathematically identical fractal dimensions may generate bosonic patterns with noticeably different scaling behavior and thus different measured fractal dimensions.

To quantify the fractal dimension of the considered bosonic patterns, we employed the box-counting method \cite{Theiler_EstimatingDimension1990}. The shape of each pattern is first extracted from a grayscale image (see, for example, Fig.~\ref{fig:patterns-driving+densities}) by applying a threshold above which points are considered part of the pattern \cite{Otsu_Threshold1979}. This thresholding procedure delineates the fractal structure within the image, effectively separating the foreground (pattern) from the background and yielding a binary representation. In this binary image, pixels above the threshold are marked as belonging to the pattern, while those below are discarded. This representation serves as the basis for further spatial analysis.

The box-counting method can be seen an inverse approach with respect to the fractal construction presented before. First, one divides the obtained binary image into a grid of $s^2$ square boxes, each with edge length $b = L/s$, where $L$ is the linear size of the system. We count the number of boxes occupied by the underlying density pattern, thereby estimating the area $A$ (in units of $b^2$) of the fractal, which corresponds to the white region shown in Fig.~\ref{fig:patterns-driving+densities}. In principle, this area scales as $A \propto s^{d_H}$, where $d_H$ is the Hausdorff (fractal) dimension. The box-counting dimension is thus given by the slope of $\log A = f(\log s)$ obtained over different scales $s$. By performing a linear fit to this curve, we extract an estimate of $d_H$, and the corresponding uncertainty is reported in the main text as the error bar on the box-counting dimension.
The patterns in Fig.~\ref{fig:patterns-driving+densities} are annotated with their respective box-counting dimensions $d_H$, which differ slightly from the theoretical nominal values $d^{th}_H$ for \textit{perfect}, infinitely self-similar fractals. For example, for the Sierpinski carpet (second column), $d^{th}_H = 1.89$, and for the Sierpinski cross (third column), $d^{th}_H = 1.46$. Nevertheless, even for these distorted and finite structures, we can determine a fractal dimension lying between 1 and~2.

For patterns with fractal dimension equal to or lower than 1 (Fig.~\ref{fig:patterns-driving+densities}, last column), we used a slightly modified version of the box-counting method. We restricted the analysis to the horizontal dimension by fixing the vertical box size to the system size $L$, while the horizontal box length still scales as $b = L/s$. Instead of square boxes of size $b^2$, we thus used rectangular boxes of size $bL$. This allows for a consistent evaluation of fractal dimensions ranging from 0 to~2. For integer-dimensional patterns (1D and 2D), we did not apply the box-counting method, as their Hausdorff dimensions are trivially $d_H = 1$ and $d_H = 2$, respectively. Nonetheless, one could apply box-counting to the patterns shown in the first and fourth columns of Fig.~\ref{fig:patterns-driving+densities}; such an analysis would yield values close to 1 or 2, but never exactly. Indeed, precisely capturing integer dimensions via box-counting is challenging due to inherent estimation errors.

At this point, we want to remind the reader explicitly that the value of the Hausdorff dimension calculated according to the equation $d_H^{\mathrm{th}} = \log(s^2 - N_{holes})/\log(s)$ and the value obtained by box-counting may differ from each other. Nevertheless, we emphasize that the fractal box-counting dimension $d_H$ is one of the key ingredients in characterizing the critical behavior of our system. For our conclusions, it is not essential whether two driving schemes share the same theoretical dimension $d_H^{\mathrm{th}}$; what matters is the Hausdorff dimension of the resulting bosonic density pattern, which is determined via box counting. In the main text, we focus on symmetric driving geometries for clarity. We expect that fractals with asymmetric driven regions will also exhibit behavior governed by their fractal dimension. To illustrate this point, we consider an asymmetric pattern constructed following the same iterative procedure as for the Sierpinski carpet, but with the upper-left square removed instead of the central one. The nominal fractal dimension is therefore identical to that of the Sierpinski carpet, since in both cases $s=3$ and $N_{\mathrm{holes}}=1$, yielding $d_H^{\mathrm{th}}=\log(9-1)/\log(3)=1.89$. However, the resulting fractal (see Fig.~\ref{fig:disconnected_pattern} (left)) is no longer symmetric with respect to the central vertical axis.

\begin{figure}
    \centering
    \includegraphics[width=0.35\linewidth]{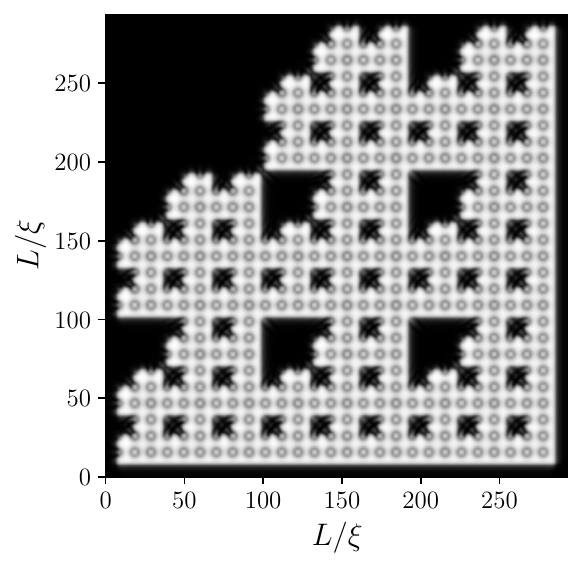}
    \includegraphics[width=0.5\linewidth]{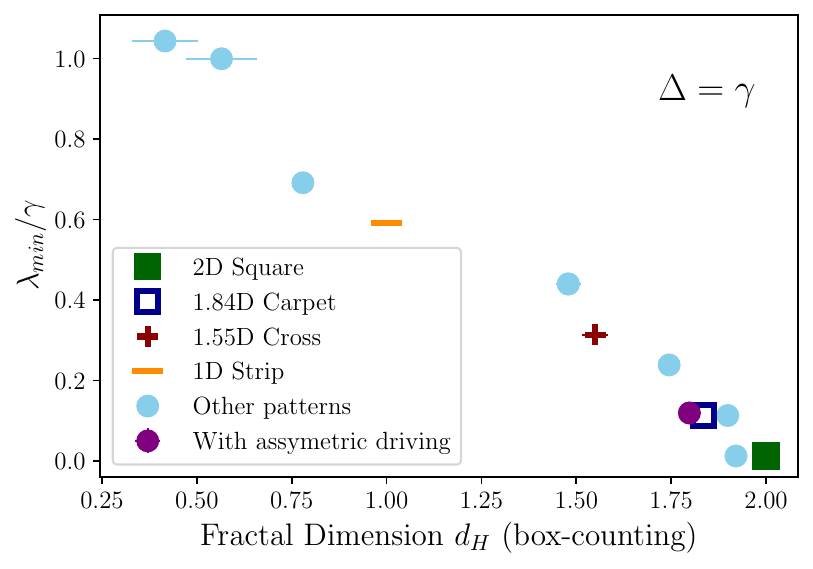}
    \caption{Left: The spatial appearance of a pattern with asymmetric driving. Its construction is the same as in Fig.~\ref{fig:dimension schema}, except now the upper left square is removed. Right: An edited version of Fig.~\ref{fig:ADR-vs-dH} of the main text where the assymetric patterns (the one shown on the left) is brought to the fore with a distinct colour (purple).}
    \label{fig:disconnected_pattern}
\end{figure}

We can test that pattern in the same way we did in the main text. We obtain the result presented on the right-hand side of Fig.~\ref{fig:disconnected_pattern}. The resulting box-counting Hausdorff dimensions $d_H$ predicts the critical dynamical behavior. For the asymmetric pattern (purple dot), we have very similar results to the Sierpinski carpet (blue empty square), which was to be expected since their nominal dimensions $d^{th}_H$ are the same.

\section{The spatial profiles of the studied fractals}
\begin{figure}[h]
    \centering
    \includegraphics[width=0.9\linewidth]{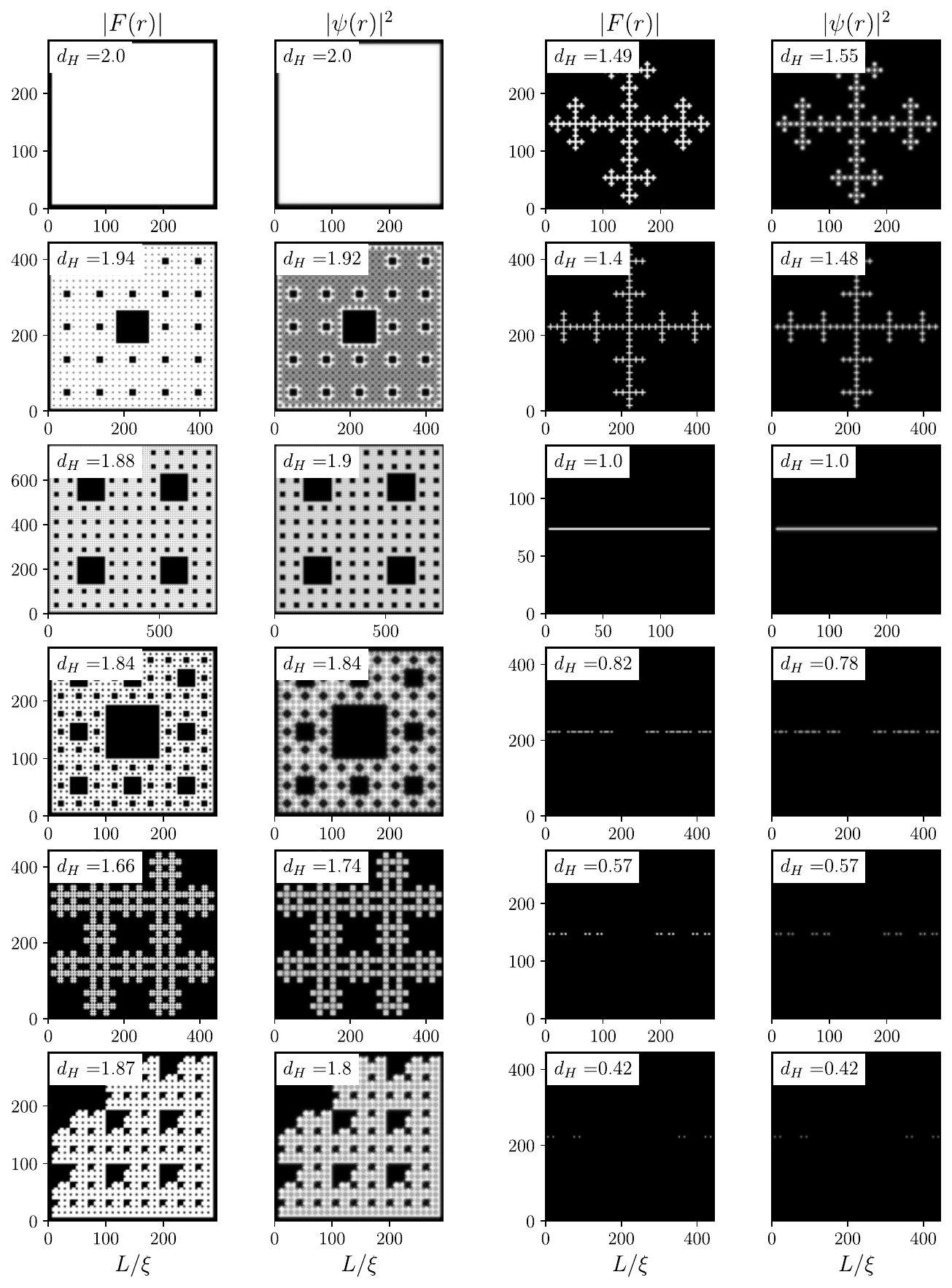}
    \caption{Left panels: 2D square geometry ($d_H = 2$) and several variations of the Sierpinski carpet with fractal dimensions in the range $1 < d_H < 2$. Right panels: Variations of the Sierpinski cross ($1 < d_H < 2$) in the first two rows, followed by the 1D strip ($d_H = 1$) in the third row. The final row shows variations of the Cantor set with $d_H < 1$. All patterns are computed with $\Delta = \gamma$, and the driving amplitudes are tuned to their respective critical points. The geometries and system sizes correspond to those used in Fig.~\ref{fig:ADR-vs-dH}.}
\label{fig:other_patterns}
\end{figure}
Figure~\ref{fig:other_patterns} shows the spatial profiles of the patterns analyzed in the main text (Fig.~\ref{fig:ADR-vs-dH}). In addition to the standard 2D and 1D reference patterns, we include several variations of the Sierpinski cross and carpet, as well as modified versions of the Cantor set with fractal dimension lower than one.

\section{Dependence of the critical driving on the detuning}

In Fig.~\ref{fig:I_min}, we extract the value of the driving intensity $I_{\min}$ such that $\lambda(I_{\min}) = \lambda_{\min}$, for a range of detuning values. As a reminder, $\lambda(I)$ denotes the asymptotic decay rate of the observed pattern as a function of the driving intensity $I$, and $\lambda_{\min}$ is its minimum value. As discussed in the main text, this minimum signals the presence of critical behavior in the system. Plotting $I_{\min}$ as a function of detuning thus allows us to trace how the onset of criticality evolves under varying external conditions.

Interestingly, the value of $I_{\min}$ closely matches the inflection point of the bistability curve derived from single-mode ($k=0$) mean-field solution for the homogeneous driving case. The inflection point corresponds to the value of the driving intensity at which the curvature of the response function (\ref{eq:bistability}) changes sign—specifically, where the second derivative of the driving intensity with respect to the density $n = |\psi|^2$ vanishes:
\begin{equation}
    \frac{d^2 I}{dn^2} = 0~.
\end{equation}
This yields the prediction
\[
I_{\min} = \frac{2\Delta}{3g}\left(\frac{\Delta^2}{9} + \frac{\gamma^2}{4}\right)~,
\]
which is shown as the grey dashed line in Fig.~\ref{fig:I_min}. The close agreement between the numerical values of $I_{\min}$ and the analytical inflection point provides a consistency check and further supports the interpretation of $\lambda_{\min}$ as a reliable marker of criticality.

\begin{figure}[t]
    \centering
    \includegraphics[width=0.6\linewidth]{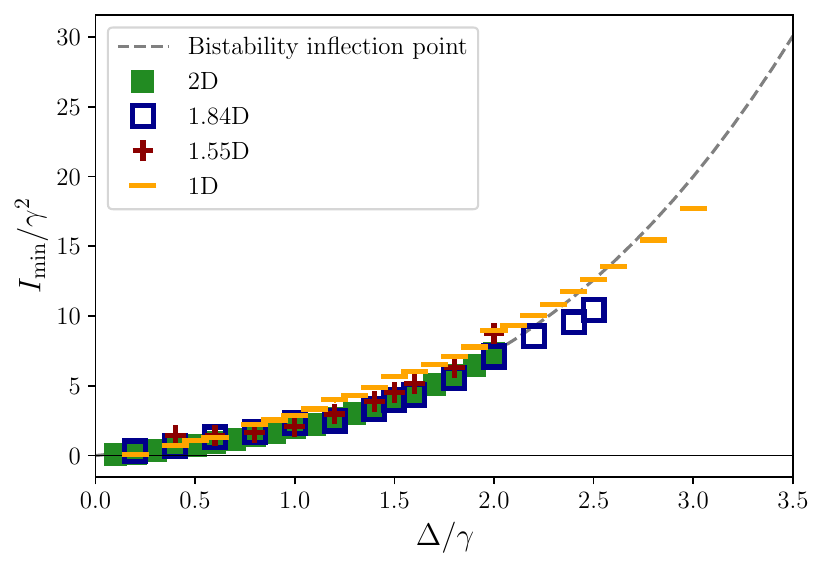}
    \caption{Evolution of the critical point $I_{\min}$ as a function of $\Delta$. A quadratic dependence is observed, consistent with estimates based on the single-mode mean-field description of the bistability curve. The inflection point of the mean-field curve (grey line) closely approximates the critical point across all geometries considered.}

    \label{fig:I_min}
\end{figure}

\section{Asymptotic decay rate and Bogoliubov instabilities}

For a homogeneous driving, we can diagonalize the problem for fluctuations on top of the steady state and obtain a spectrum of the so-called dissipative Bogoliubov modes \cite{RMP2013}. By analyzing the dispersion relation given in equation~(\ref{eq:bogo}), we can interpret the phase transition as the emergence of an unstable mode.

\begin{equation} 
    \omega_{\mathrm{Bog}}(\mathbf{k})= \pm\left[\left(-\Delta+\frac{\hbar \mathbf{k}^2}{2 m}+2 gn\right)^2-\left(g n\right)^2\right]^{1 / 2} -i\frac{\gamma}{2}~.
\label{eq:bogo}
\end{equation}

All modes decay with the natural polariton decay rate $\gamma$; however, this rate can be reduced and is determined by the imaginary part of $\omega_{\mathrm{Bog}}$. If $\Im\left(\omega_{\mathrm{Bog}}(\mathbf{k})\right) < 0$, the mode is stable and has a linewidth given by $2\Im\left(\omega_{\mathrm{Bog}}(\mathbf{k})\right)$. The factor of 2 arises because we are considering the evolution of the bosonic density, $n(t) = |\psi(t)|^2$. 

An unstable mode corresponds to $\Im\left(\omega_{\mathrm{Bog}}(\mathbf{k})\right) \geq 0$. In this case, the linewidth vanishes, the asymptotic decay rate tends to zero, and we can interpret the system as undergoing critical behavior.

We can define the minimal linewidth $\lambda_{\mathrm{Bog}}$ associated with the temporal evolution of the density of a homogeneous phase with steady-state density $n_{ss}$ as:
\begin{equation}
\lambda_{\mathrm{Bog}}= 
  \left\{
    \begin{array}{l}
      2|\max_{k,n_{ss}}\Im\left(\omega_{\mathrm{Bog}}(k,n_{ss})\right)| \text{ if stable}\\
      0 \text{ else  .}
    \end{array}
  \right.
\end{equation}

After calculations, this leads to the following expression:
\begin{equation}
\lambda_{\mathrm{Bog}}= 
  \left\{
    \begin{array}{l}
    \gamma  \text{   if }\Delta<0 \\
      \gamma-\frac{2}{\sqrt{3}}\Delta \text{   if }0<\Delta<\frac{\sqrt{3}}{2}\gamma \\
      0 \text{   else . }
    \end{array}
  \right.
\end{equation}

This dependence is shown as the dotted line in Fig.~\ref{fig:bogo instabilities}, alongside numerical results for the 2D square geometry. Both the analytical predictions and the numerical data display a step-like behavior with a linear slope between the upper and lower values. The two curves coincide for $\Delta \gtrsim 1.0\gamma$, where the system becomes critical and $\lambda_{\min} = 0$. Deviations at other detuning values can be primarily attributed to finite-size effects.

\begin{figure}[h]
    \centering
    \includegraphics[width=0.6\linewidth]{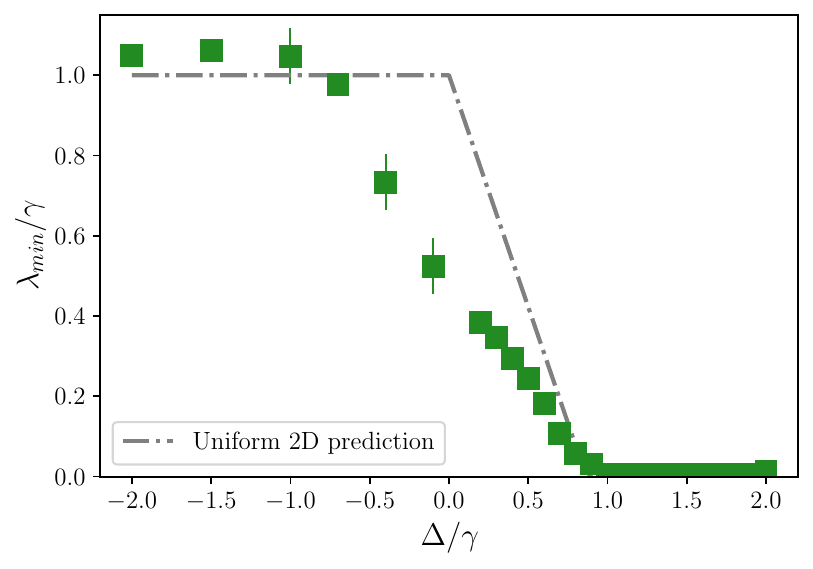}
    \caption{Minimal asymptotic decay rate for the 2D square geometry (green squares), computed for a system size $L = 81l$. Vertical error bars indicate the difference in $\lambda_{\min}$ between this size and a smaller system with $L = 27l$. The grey dash-dotted line shows the linewidth $\lambda_{\mathrm{Bog}}$ predicted by Bogoliubov analysis for a uniform 2D system. The two curves exhibit similar behavior, with discrepancies attributed to finite-size effects.}

    \label{fig:bogo instabilities}
\end{figure}

\section{Time evolution of the density profile}

Additional insight can be gained by examining the time evolution of the density profile. In particular, we focus on the spatial structure of the slowest decaying mode in the fractal geometry corresponding to the Sierpinski carpet ($d_H \approx 1.8$). Figure~\ref{fig:spatial-decay} displays the spatial dependence of the density deviation from the steady state at different time steps; specifically, we plot $|n_{ss}(\textbf{r}) - n(\textbf{r}, t)|$ for various times $t$. This local information can be compared with the global behavior shown in the upper panel of Fig.~\ref{fig:spatial-decay}, where the evolution of the spatially averaged density $\bar{n}(t)$ is presented for the same system.

\begin{figure}[t]
    \centering
    \includegraphics[width=0.85\linewidth]{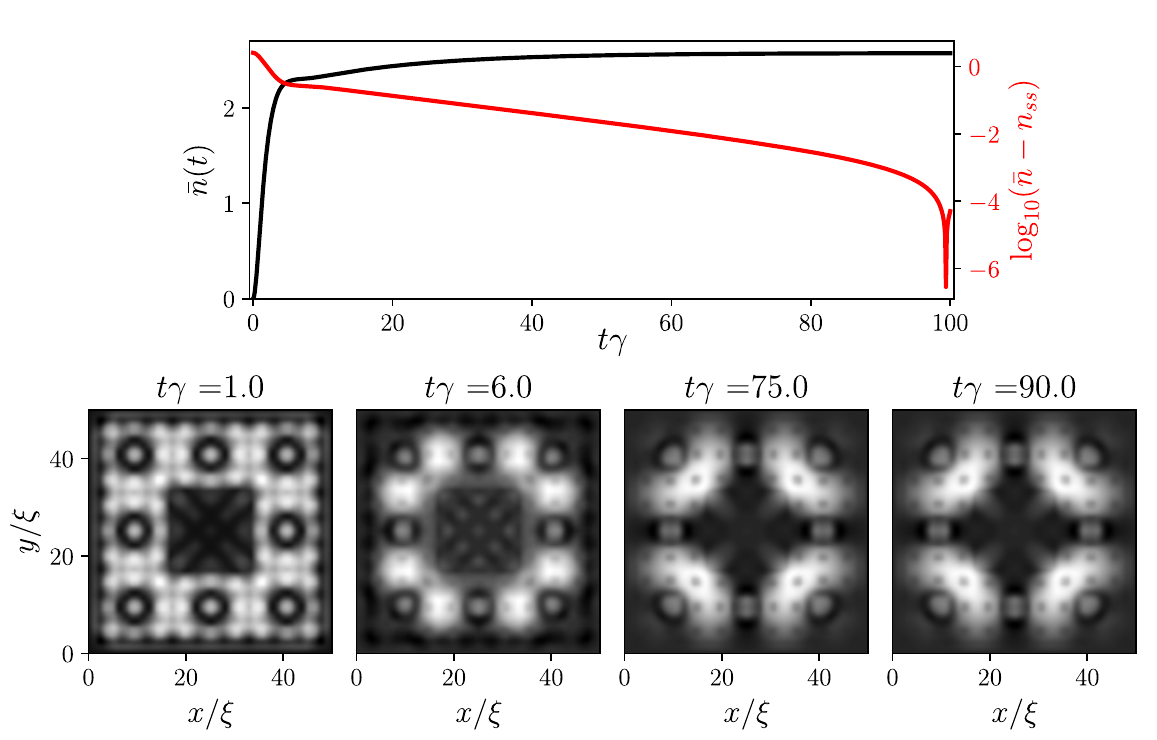}
    \caption{Lower panels: Spatial profile of the density deviation from the steady state at different time steps $t$, given by $|n_{ss}(\textbf{r}) - n(\textbf{r}, t)|$, for the Sierpinski carpet at the crossover point $I = 2.3\gamma^2$ with $\Delta = \gamma$. Upper panel: Time evolution of the spatially averaged density $\bar{n}(t) = \int \mathrm{d}^2x\, n(t, x)$. In red, we plot the logarithm of the deviation from the steady state, making the asymptotic decay rate clearly visible.}

    \label{fig:spatial-decay}
\end{figure}

At early times, the density $n(\textbf{r}, t)$ is close to zero throughout the system, so the difference shown in the first lower panel of Fig.~\ref{fig:spatial-decay} primarily reflects the shape of the final steady state. At later times, as shown in the final panels, a distinct pattern begins to emerge. This structure corresponds to the longest-lived mode—i.e., the one governing the asymptotic decay rate. In the Liouvillian framework, this is the eigenstate of the Lindbladian superoperator associated with the Liouvillian gap (the smallest nonzero eigenvalue). Thus, we are observing the mode that is key to characterizing the system’s critical behavior.

For a homogeneous condensate, this connects to the dispersion relation $\omega(\textbf{k})$ given in Eq.~(\ref{eq:bogo}). In certain cases, the slowest-decaying mode (i.e., the one with the largest imaginary part of $\omega$) occurs at a nonzero wavevector $\textbf{k} \neq \vec{0}$, meaning that the longest-lived mode may not share the spatial symmetry of the driving field. In our case, the resulting mode exhibits a distinctive and nontrivial spatial structure, which will be the subject of future investigation.

\section{Critical Detuning Independence from the interaction strength within the Non-Equilibrium Gross-Pitaevskii Framework}
\label{sec:g}

Within the mean-field treatment employed in this work, changing the nonlinear interaction constant $g$ does not alter the underlying physics. In other words, when analyzing the Gross-Pitaevskii equation~(\ref{eq:GPE}), varying $g$ does not affect the qualitative or quantitative results.

To illustrate this, consider two systems with nonlinear interaction strengths $g_1$ and $g_2$, and assume identical detuning and dissipation parameters: $\Delta_1 = \Delta_2 = \Delta$ and $\gamma_1 = \gamma_2 = \gamma$. Let us express $g_2 = g_1 / \eta^2$, where $\eta$ is a real scaling factor. The dynamics of system 2 is then governed by:
\begin{equation}
    i\frac{d\psi_2}{dt} = \left[-\Delta - \frac{\hbar \nabla^2}{2m} + \frac{g_1}{\eta^2} |\psi_2|^2 - i\frac{\gamma}{2} \right]\psi_2 + iF_2(\textbf{r})~.
\end{equation}
If we rescale the driving field and wavefunction as $F_2 = \eta F_1$ and $\psi_2 = \eta \psi_1$, the equation for $\psi_2$ becomes identical to that of system 1. This rescaling is permissible because both $F$ and $\psi$ are degrees of freedom of the system and, in practice, are often expressed in arbitrary units in experimental setups.

Therefore, as long as the mean-field approximation holds, our results are invariant under changes in the absolute value of $g$. However, this invariance is not expected to persist in the strongly interacting regime, which lies beyond the scope of the present work and will be explored in future studies.

\end{document}